\documentclass[twoside,twocolumn,english,aps,showpacs,prl,superscriptaddress]{revtex4-1}
\usepackage[T1]{fontenc}
\usepackage[latin9]{inputenc}
\usepackage{color}
\usepackage{bm}
\usepackage{amsmath}
\usepackage{amssymb}
\usepackage{graphicx}
\usepackage{esint}
\usepackage{mathrsfs}

\makeatletter

\makeatletter
\newcommand*{\rom}[1]{\expandafter\@slowromancap\romannumeral #1@}
\makeatother

\usepackage{times}{\Large }

\makeatother

\usepackage{babel}
\begin{document}

\title{Reveal the non-local coherent nature from a dissipative Majorana teleportation}

\author{Donghao Liu}

\affiliation{State Key Laboratory of Low Dimensional Quantum Physics, Department of Physics, Tsinghua University, Beijing, 100084, China}

\author{Zhan Cao}
\affiliation{Beijing Academy of Quantum Information Sciences, Beijing 100193, China}
\affiliation{State Key Laboratory of Low Dimensional Quantum Physics, Department of Physics, Tsinghua University, Beijing, 100084, China}

\author{Hao Zhang}
\affiliation{State Key Laboratory of Low Dimensional Quantum Physics, Department of Physics, Tsinghua University, Beijing, 100084, China}
\affiliation{Beijing Academy of Quantum Information Sciences, Beijing 100193, China}
\affiliation{Frontier Science Center for Quantum Information, Beijing 100184, China}

\author{Dong E. Liu}
\email{Corresponding to: dongeliu@mail.tsinghua.edu.cn}
\affiliation{State Key Laboratory of Low Dimensional Quantum Physics, Department of Physics, Tsinghua University, Beijing, 100084, China}
\affiliation{Frontier Science Center for Quantum Information, Beijing 100184, China}

\begin{abstract}
The recent observations of Majorana resonance increase our confidence in the realization of the first topological qubit.
The non-local coherent nature of Majorana devices is the key factor for the topological protection of these qubits from decoherence. Direct observation of this coherent nature could provide a first-step benchmarking scheme to validate Majorana qubit quality, which could be the next experimental milestone.  We propose a simple transport scheme with a Majorana island device along with a dissipative environment in the electrodes. We find that the dissipative environment renormalizes the quantum transport in significantly different ways: As the temperature is reduced, while the conductance for Majorana coherent teleportation increases, all other incoherent signals are strongly suppressed due to dissipation. This unique conductance scaling behavior is a clear benchmark to reveal the non-local coherent nature of Majorana devices.
\end{abstract}


\date{\today}

\maketitle


{\em Introduction--}. The realization of Majorana zero modes (MZMs)~\cite{ReadGreen,1DwiresKitaev} provides a promising platform to study novel fundamental physics, e.g. non-Abelian braiding statistics~\cite{NonAbelian77,NonAbelian89,Ivanov2001NonAbelian,TQCreview}, and has potential applications in quantum information processing and topological quantum computation~\cite{kitaev,TQCreview}. Many proposals for realizing MZMs in topological superconductors (SCs) have been put forward ~\cite{Fu&Kane08,SatoPRL09,Sau10,LutchynPRL10,1DwiresOreg,Sau10,Alicea10,CookPRB'11,AliceaRev}, and lead to recent experimental progress in the realization and detection of MZM in both one-dimensional~\cite{Mourik2012,Rokhinson2012,Deng2012,Das2012,Churchill2013,Finck2013,Nadj-Perge14,Albrecht16,deng2016Majorana,Zhang2017Ballistic,ZhangNN2018,ZhangNature2018} and  two-dimensional platforms~\cite{Sun2016Majorana,Wang2018Evidence,TamegaiVortexMZM2019,Liu2018Robust,Fornieri2019NatureTJJ,Ren2019Nature,Luli2019MZM}. The quantized Majorana conductance at $2e^2/h$, observed in nanowire devices~\cite{ZhangNature2018}, closes one chapter in tunneling spectroscopy based on the simplest device set-ups. Additionally, some clues of Majorana conductance plateau were also shown in the vortex core of topological SC~\cite{MajoranaPlateauHD,02FengDLQuantizedM2019}. Those experimental observations are gearing up for next-step discoveries~\cite{ZhangLiuReview} towards the realization of non-Abelian braiding experiments and Majorana qubits.

So far, experimental activities of Majorana research mainly focus on the Majorana resonance behavior or its oscillating splittings~\cite{Albrecht16}. 
These experiments, aiming at verifying the existence of Majorana bound states, reveal little information on the non-local coherence nature between two Majorana pairs. It is this non-local coherent property which enables the non-local information storage and makes Majorana qubit robust against decoherence.
Therefore experimentally demonstrating the Majorana non-local coherence would be a milestone towards the final braiding experiment and topological qubits. 
Among the earlier proposed schemes, a well-studied one is the fractional Josephson effect~\cite{1DwiresKitaev,Kwon}, which has undergone a lot of experimental activities~\cite{Rokhinson2012,laroche2019Josephson}. However, the contamination mechanisms of the $4\pi$ periodicity mix both incoherent quasi-particle events~\cite{Fu&Kane09,LutchynPRL10} and dynamical Landau-Zener processes~\cite{houzet2013dynamics,badiane2013ac}. Therefore, the deviation of $4\pi$ periodicity may not be a valuable benchmark for Majorana device coherence. Another important scheme is the non-local coherent teleportation proposed by Fu~\cite{MajoranaTeleportation} for a floating Majorana superconducting island with finite charging energy. The electron transport through the island is mediated by the teleportation process: an electron injected into the MZM at one end can tunnel out from the other MZM at the other end while still maintaining its phase coherence even for a long wire-distance. In order to benchmark the coherent nature, an Aharonov-Bohm (AB) interference experiment~\cite{MajoranaTeleportation} is thought to be necessary, where the electron coherence in a parallel normal channel needs also to be maintained in a loop-type device. Those requirements greatly increase the complexity and difficulty in experiments~\cite{Nature2017Network}. Therefore, a realistic scheme with more experimentally feasible conditions and sharp transport signals is preferable to reveal the Majorana non-local coherence.

In this letter, we propose such an experimentally feasible scheme based on a Majorana island device coupled to a dissipative bath. The dissipative bath can be realized by making part of the leads highly resistive (comparable to von-Klitzing resistance $h/e^2$)~\cite{Nazarov1992charge,Dong2012Nature,MebrahtuNaturePhy}. In our scheme, we only require a simple conductance measurement to reveal the coherence nature without the AB loop structure. The intuitive trust of this scheme is from the fact that the dissipative environments suppress all low-energy transport signals except for the symmetric coherent resonant level system~\cite{Dong2012Nature,DongPRB2014}. We map our system to a dissipative resonant level model~\cite{Dong2012Nature,DongPRB2014}, and show that, for symmetric lead-island couplings, the conductance from coherent Majorana teleportation increases as the temperature is reduced. While, for all other non-Majorana regimes where incoherent transports dominate, the conductance with the same parameter conditions show a power-law decay as the temperature is reduced.  In addition, we want to point out that the dissipative electrode lead (which could help us to reveal the coherence or decoherence) in our scheme is quite different from the disorder~\cite{BrouwerDisorder2011,SauDisorder2012,JieLiuDisorder2012,TakeiSoftGap,HuiDisorder2015,ColeDisorder2016,DEL-Disorder2018,ChangNatureNano,KouwenhovenNanoLett} in the Majorana island, which can cause in-gap soft modes and contaminate their coherence. Therefore, the conductance enhancement behavior in the dissipative environment can serve as a clear benchmark for the coherent nature of Majorana devices.

\begin{figure}
\includegraphics[width=1\columnwidth]{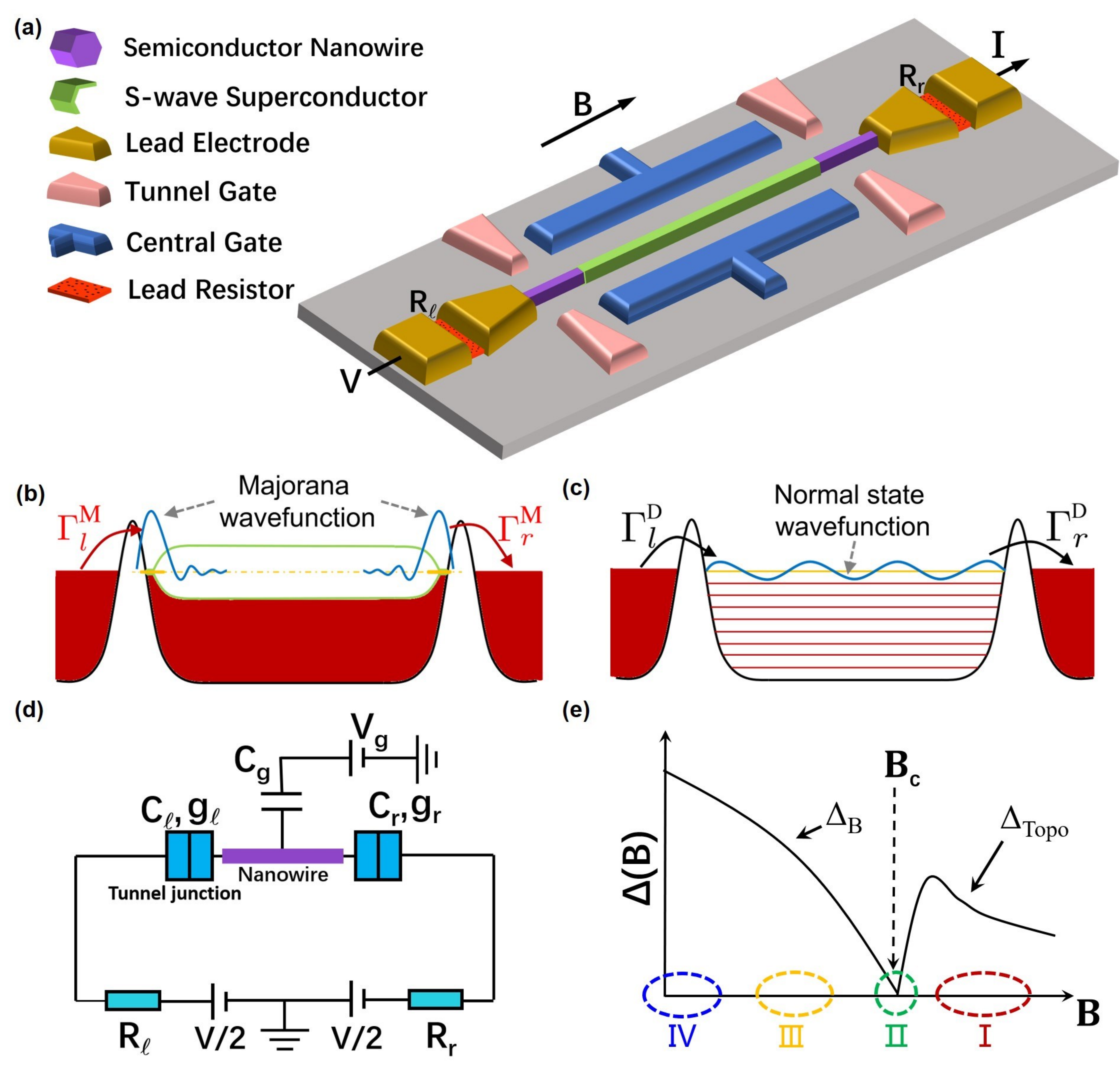}

\caption{\label{Fig01} (a) Illustration of the system set-up. (b),(c) Illustration of Majorana teleportation and normal state sequential tunneling. (d) An equivalent circuit diagram of (a). The nanowire/superconductor island is connected to the leads via two tunneling junctions characterized by capacitance $C_{l,r}$ and dimensionless conductance $g_{l,r}$ (conductance $G_{l,r}=g_{l,r}2e^2/h$). (e) Phase diagram of the proximitized nanowire. The quasiparticle gaps in the trivial and topological phase are labeled by $\Delta_\textit{B}$ and $\Delta_\textit{Topo}$, respectively.}
\end{figure}

{\em Model--} The system set-up is shown in Fig. 1(a). A semiconductor nanowire with Rashba spin-orbit coupling is in proximity with an s-wave superconductor. A magnetic field B is applied in parallel with the nanowire to realize a tunable topological SC phase~\cite{LutchynPRL10,1DwiresOreg}. The central gates control the nanowire chemical potential, while the tunnel-gates control the coupling between the (left and right) leads and the proximitized nanowire. The system also couples to a dissipative bath realized by adding on-chip resistors in source and drain leads~\cite{Dong2012Nature}. This can be achieved by replacing part of the electrode (typically Au, 100 nm thick: brown in Fig. \ref{Fig01}(a)) with a thin (around 10 nm thick: red in Fig. \ref{Fig01}(a)) long resistive metal strip (e.g. Cr). For the dissipation effect to be effective, the resistive part needs to be on the device chip (at low temperature) and close the electrodes which form good Ohmic contacts with the nanowire. An equivalent circuit diagram is shown in Fig. \ref{Fig01}(d) including a familiar quantum (lead-island) part coupled to a classical R-C circuit.

The central nanowire island is electrically floated such that electrons in the island feel an electrostatic energy:  $U_{N}=E_{c}\left(N-V_{g}C_g/e\right)^{2}$, where  $E_{c}=e^{2}/2C_{\Sigma}$ is the island charging energy and $C_{\Sigma}=C_l+C_r+C_g$ is the total capacitance. $N$ indicates the total excess electron number of the island. Here, $C_g$ ($C_l$, $C_r$) is the central gate (left, right junction) capacitance, and $V_{g}$ is the voltage applied on the central gate. The nanowire shows induced superconductivity with an energy gap $\Delta_\textit{B}$ at the magnetic
field $B$ as shown in Fig. \ref{Fig01}(e). The gap closes at critical $B_{c}$ indicating a topological phase transition from a trivial phase to a topological phase with MZMs at the island ends. The Hamiltonian of the island is
\begin{equation}
H_{\textit{wire}}=\sum_{\alpha}\epsilon_{\alpha}\gamma_{\alpha}^{\dagger}\gamma_{\alpha}+U_{N}.\label{eq:nanowire Hamiltonian}
\end{equation}
Here,  $\gamma_{\alpha}$ is the Bogoliubov quasiparticle operator and $\epsilon_{\alpha}$ is the quasiparticle energy. In trivial phases ($B<B_{c}$), the energy levels above the gap can be approximately expressed as $\epsilon_{\alpha}=\sqrt{\Delta_\textit{B}^{2}+\xi_{\alpha}^{2}}$ with $\xi_{\alpha}$ being the normal state energy. In topological SC phase ($B>B_{c}$), there are two-fold degenerate ground states separated from the continuum by a topological gap  $\Delta_\textit{Topo}$.

Including the left and right leads along with dissipative environment (resistor $R_l$ and $R_r$ in two leads), the total Hamiltonian of the system can be written as
\begin{equation}
H_{\textit{tot}}=H_{\textit{leads}}+H_{\textit{wire}}+H_{\textit{T}}+H_\textit{env},
\end{equation}
where $H_{\textit{leads}}=\sum_{j,p\sigma}\xi_{j,p}c_{j,p\sigma}^{\dagger}c_{j,p\sigma}$ describes the free electrons in the left ($j=l$) and right ($j=r$) leads with electron energy $\xi_{p}$ at orbital $p$.
\begin{equation}
H_{\textit{T}}=\sum_{j,\sigma}t_{j}d_{\sigma}^{\dagger}\left(\bold{r}_j\right)c_{\sigma}\left(\bold{r}_j\right)e^{-i\varphi_{j}}+\textit{H.c.},\label{eq:tunneling Hamiltonian}
\end{equation}
which describes the tunneling between the nanowire and the leads. $d_{\sigma}\left(\bold{r}_j\right)$ and $c_{\sigma}\left(\bold{r}_j\right)$ are the electron field operator in the nanowire and lead separately, with spin $\sigma$ and evaluated at the specific position of the junction-$j$~\cite{supp}. $t_{j}$ is the average tunneling amplitude at that junction. The tunneling across that single junction gives rise to the conductance $G_j=g_{j}(2e^2/h)$ with $g_j\propto\left|t_{j}\right|^{2}\nu_{j}\nu$ where $\nu_{j}$ ($\nu$) is the density of state (DOS) at the leads (nanowire). Later, we will require symmetric island-lead couplings $g_l=g_r\equiv g$ by fine tuning a single parameter. $H_\textit{env}$ describes the dissipative environment, and can be modeled by the classical circuit~\cite{Nazarov1992charge}. To incorporate the dissipation effect in our system, we add a phase operator $e^{-i\varphi_{j}}$ in the tunneling Hamiltonian~\cite{Nazarov1992charge,DongPRB2014}; and this operator couples the quantum tunneling with the classical circuit. This phase operator change the total charge of the junction capacitor by one electron (note the charge-phase conjugation $\left[\varphi_{j},Q_{j'}\right]=ie\delta_{jj'}$ where $Q_{j}$ is the charge of the junction capacitor). For convenience of calculations, we consider small gate capacitance $C_g\ll C_{l,r}$, then the only relevant phase is $\varphi=\varphi_l-\varphi_r$ which corresponds to the charge transfer between the left and right junction; and for general cases, the physics will not change qualitatively~\cite{DongPRB2014}. The correlation function of this phase operator can be obtained from the corresponding R-C circuit, and results in $\left\langle e^{i\varphi(t)}e^{-i\varphi(0)}\right\rangle\sim (\omega_R t)^{-2r}$ where $r=R/R_K$ with von-Klitzing resistance $R_K=h/e^2$, and $\omega_R=(R C)^{-1}$ with $R=R_l+R_r$ and $C=C_l C_r/(C_l+C_r)$.

{\em Coherent teleportation vs incoherent transport--}. The electron transports through a confined island with charging energy show Coulomb blockade (CB) oscillations~\cite{CB-Review,bruus2004book}. For a trivial SC island, the low-energy events correspond to Cooper-pair transport with $2e-$ periodicity in CB oscillation~\cite{hekking1993CBinSC,van2016PRB}. However, the presence of MZMs in a topological SC island causes a regular single electron tunneling with $1e-$ periodicity, which is the same as the periodicity for a normal metal island. Why is the Majorana teleportation special? In fact, the key underlying physics is quite unique. For Majorana teleportation~\cite{MajoranaTeleportation} as shown in Fig.~\ref{Fig01} (a), the Majorana wave-function is localized at the wire's two ends, and overlaps strongly with electrode leads. The lead-Majorana coupling strength is represented by $\Gamma^\textit{M}\sim g \Delta_\textit{Topo}$ which is the reverse electron lifetime on the island electron. Then, for $T<\Gamma^\textit{M}$ and gate-voltage tuned close to CB peaks, the electron can tunnel into one Majorana state and out of the other one and maintain its phase coherence. On the other hand, the normal state wavefunction widely spreads out over the wire as shown in Fig.~\ref{Fig01}(c). In order to reach the other side, the electrons have to propagate through the whole wire with a typical time $\hbar/\Gamma^{D}$ (note $\Gamma^{D}\sim g_j \delta$ with a tiny level spacing $\delta$). Therefore, in the realistic regime, $g\delta \ll T \lesssim g \Delta_\textit{Topo}$ ( $\delta\ll\Delta_\textit{Topo}$ which requires a long island), only Majorana teleportation maintains phase coherence; and on the contrary, all other normal states generate incoherent transport near CB peak. Although the regular conductance measurement cannot tell the difference: both show 1e-periodic CBs, we will see the dissipative environment could help us to reveal the coherent signature without a more sophisticated interferometer set-up.

\begin{figure}
\includegraphics[width=1\columnwidth]{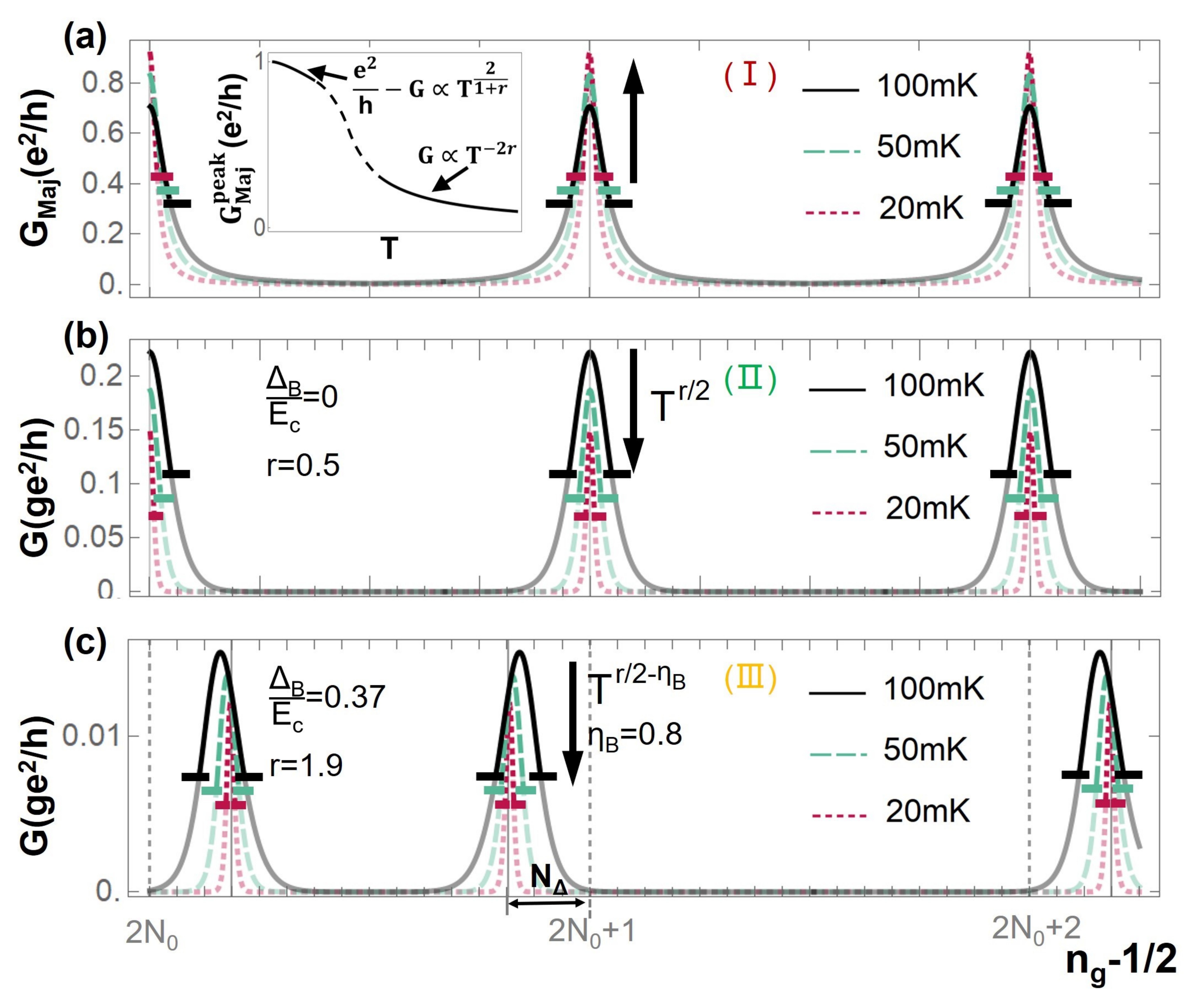}
\caption{\label{G_Single} The dissipative CB conductance as a function of dimensionless central gate potential $n_g\equiv V_g C_g /e$ for different temperature in region ({\rom{1}})-({\rom{3}}) of Fig.~{\ref{Fig01}(e)}. (a) A schematic drawing of the dissipative CB conductance in Majorana phase (region ({\rom{1}})) by fitting the results in~\cite{Dong2012Nature,HaroldMaj2016} with $T\ll\Gamma^\textit{M},\omega_R$. The inset shows the peak conductance as a function of $T$ in a larger range. (b) The dissipative CB conductance in normal metal phase (region (\rom{2})) with $\Delta_\textit{B}<\delta$ and $E_{c}=2.32K$. (c) The dissipative CB conductance of a superconductor in region ({\rom{3}}) where $T\ll\Delta_\textit{B}<E_{c}$ with $\Delta_\textit{B}=0.86K$, $E_{c}=2.32K$. The peak positions shift about $N_{\Delta}$ with $N_{\Delta}\equiv \Delta/\left(2E_c\right)$. $N_0$ represents an integer.}
\end{figure}

{\em Dissipative Majorana teleportation--}. As the nanowire enters into the topological region (\rom{1}) shown in Fig.~\ref{Fig01}(e), the transport is dominated by the resonant tunneling through the zero energy Majorana state. With dissipation, this system can be mapped to a dissipative resonant level model~\cite{Dong2012Nature,DongPRB2014}. For symmetric island-lead couplings $g_l=g_r$, the dissipation effect cannot qualitatively change the low-energy transport behavior for an originally coherent resonant tunneling. For our model, as long as $r=R/R_K<2$ for arbitrary $\Gamma^\textit{M}$ or $2\leq r<3$ for $\Gamma^\textit{M}$ above a critical value,  the CB peak height still increases as the temperature is reduced~\cite{Dong2012Nature,DongPRB2014} as shown in Fig.~\ref{G_Single} (a). Dissipation only slightly changes the scaling function: At low temperature $T\ll\Gamma^\textit{M},\omega_R$, the peak height $G_{\textit{M}}^{\textit{peak}}$ reaches $e^2/h$ as $e^2/h-G_{\textit{M}}^{\textit{peak}}\propto T^{{2}/{(1+r)}}$~\cite{HaroldMaj2016}; while at higher $T$, the peak height increases as $T^{-2r}$ when $T$ decreases~\cite{DongPRB2014}. On the other hand, we will see dissipation significantly renormalizes the low-energy transport, and results in an opposite temperature dependence for other processes, which are incoherent, in the next few sections.

{\em Dissipative sequential tunneling--}. As mentioned earlier, in the non-topological regime, the electron transport near CB peaks through the nanowire loses the phase information. Those tunneling processes can be treated sequentially through the two junctions; and therefore, the conductance can be obtained from a rate equation method~\cite{beenakker1991theory,glazman2005review}. Including the dissipation effect, we derive the general form of the zero-bias conductance near the CB peaks (see SI~\cite{supp} for more details):
\begin{align}
  G=&\frac{\Lambda e^{2}}{h}\frac{ g_{l}g_{r}}{g_{l}+g_{r}}\frac{\delta}{ T}W_{\mathrm{eq}}\left(N\right)\sum_{\alpha}\!\!\int\!\! d\xi_p\!\! \int\!\! dE\! \left[1-F_{\mathrm{eq}}\left(\epsilon_{\alpha}|N\right)\right]   \nonumber\\
  &\times f\left(\xi_p\right)\delta\left(\epsilon_{\alpha}-\xi_{p}+U_{N+1}-U_{N}+E\right)P\left(E\right),
  \label{eq:general single electron CB peak}
\end{align}
which is from the processes that a lead-electron with energy $\xi_p$  jumps into an unoccupied level $\alpha$ with energy $\epsilon$$_{\alpha}$ in the island. Note that $\Lambda\sim 1$ is a dimensionless parameter associated to the SC wavefunction~\cite{supp}. The energy conservation requires $\epsilon_{\alpha}+U_{N+1}+E=U_{N}+\xi_{p}$ where the energy exchange $E$ with environment is allowed. The probability function $P\left(E\right)$ of such an exchange can be written as the Fourier transform of the phase-phase correlation function~\cite{Nazarov1992charge,supp}
\begin{align}
P\left(E\right)&=\frac{1}{2\pi\hbar}\int_{-\infty}^{+\infty}\mathrm{d}t\left\langle e^{i\varphi(t)}e^{-i\varphi(0)}\right\rangle e^{\frac{i}{\hbar}Et}\nonumber\\
&\begin{array}{cc} \propto T^{\frac{r}{2}-1}e^{\frac{E}{2T}}\left|\mathit{\Gamma}\left(\frac{r}{4}+i\frac{E}{2\pi T}\right)\right|^{2}
 & ,\text{for}\; T\ll\omega_{R}\end{array},\label{eq:PE}
\end{align}
where $\mathit{\Gamma}\left(z\right)$ is the gamma function. $W_{\mathrm{eq}}\left(N\right)$ is the probability that the isolated island contains $N$ electrons in equilibrium, $F_{\mathrm{eq}}\left(\epsilon_{\alpha}|N\right)$ is the equilibrium conditional probability that level $\alpha$ is occupied given the island electron number $N$, and $f(x)$ is the standard Fermi-Dirac distribution~\cite{supp}. Next, we will show how the dissipation affects the conductance in different  non-topological regimes ($B<B_{c}$), respectively.

Near the SC gap closing point in region (\rom{2}) of Fig.\ref{Fig01}(e), the nanowire corresponds to a metallic phase and behaves like a normal metal island. For a long nanowire, the level spacing is small such that  $\delta\ll T$, the discrete energy spectrum of the island can be treated as a continuum. In this limit, $W_{\mathrm{eq}}\left(N\right)\approx1/\left[1+e^{-\left(U_{N+1}-U_{N}\right)/T}\right]$ and $F_{\mathrm{eq}}\left(\epsilon_{\alpha}|N\right)$ can be approximated as Fermi-Dirac distribution $f\left(\epsilon_{\alpha}\right)$~\cite{supp}. Applying the integration in Eq. (\ref{eq:general single electron CB peak}), we obtain the conductance near CB peaks with peak height proportional to $T^{r/2}$~\cite{supp}.  The CB conductance oscillations with $1e$-periodicity for different temperatures is shown in Fig.~\ref{G_Single}(b), where the dissipation parameter is $r=0.5$. Different from the coherent Majorana teleportation, the dissipation effect significantly suppresses the conductance at low energy. Note that this calculation is valid only near the CB peaks; and beyond the peak width (labeled by horizontal bars in the figures), the co-tunneling events need to be included in the calculation.

Secondly, we consider the case that the system has a finite but smaller gap $\Delta_\textit{B}<E_{c}$ (region (\rom{3})). In this case, adding an unpaired electron is still energetically preferred, so the single electron tunneling contributes to the conductance~\cite{Nazarov1992single,van2016PRB}. The degeneracy happens at $U_{N}\approx U_{N+1}\pm\Delta_\textit{B}$ ("$+$" for even $N$ and "$-$" for odd $N$); therefore, the position of the peaks will shift to the right (left) by about $N_{\Delta}$ for even (odd) $N$ as shown in Fig.~\ref{G_Single}(c). In the conductance calculation, we need to include the tunneling processes where electrons enter into the levels above the SC gap in the island, and thus we have  $W_{\mathrm{eq}}\left(N\right)\approx 1/\left[1+\sum_{\alpha}e^{-\left(\epsilon_{\alpha}+U_{N+1}-U_{N}\right)/T}\right]$ near the charge degenerate point. In region (\rom{3}), the SC gap is still big enough ($\Delta_\textit{B}\gg T$) to prohibit thermal excitations of quasi-particle states, so $\left[1-F_{\mathrm{eq}}\left(\epsilon_{\alpha}|N\right)\right]\approx 1$.
Then following Eq.(\ref{eq:general single electron CB peak}) we can obtain the conductance near the CB peaks~\cite{supp}. The conductance peak height shows a special temperature scaling $T^{r/2-\eta_B}$, where an extra scaling factor $\eta_B$ tends to one for $\Delta_\textit{B}\gg T$ and zero at the gap closing point (refer to SI~\cite{supp} for more details). The dissipative conductance at $r=1.9$ and $\Delta_B=0.86K$ is shown in Fig.~\ref{G_Single}(c). We can see that the dissipation still renormalizes the low energy conductance in quite a different way compared with Majorana teleportation.


\begin{figure}
\centering
\includegraphics[width=1\columnwidth]{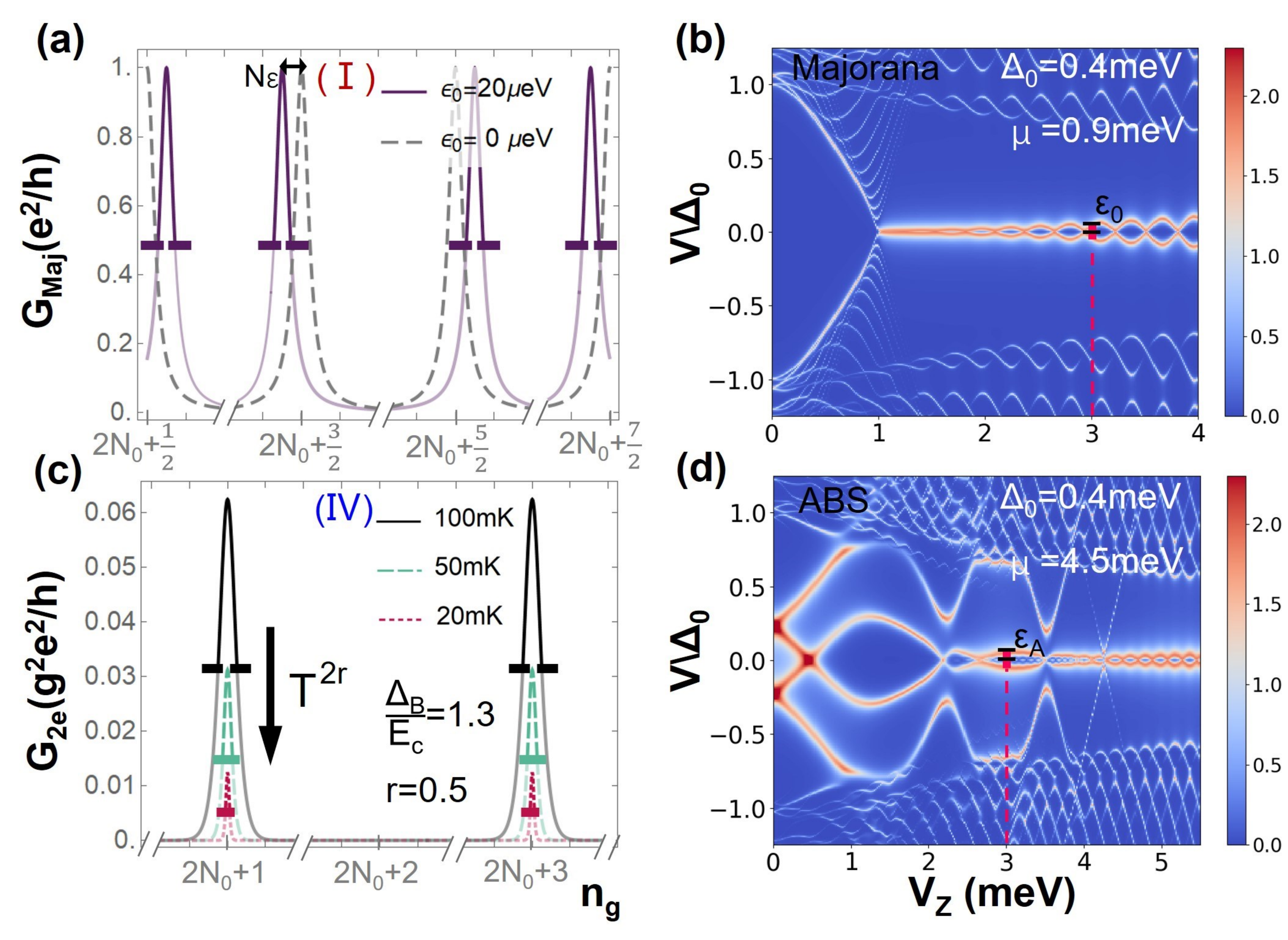}
\caption{\label{G_Splitting} (a) The Majorana CB conductance when there is a splitting $\epsilon_0$ as shown in (b). The solid and dashed lines represent the conductance with and without splitting respectively. The peak position will be shifted by $N_{\epsilon}\equiv\epsilon_0/\left(2E_c\right)$ when splitting occurs. (c) The dissipative CB conductance of a superconductor in region (\rom{4}) of Fig.~{\ref{Fig01}}(e) where $\Delta_\textit{B}>E_{c}$ with $\Delta_\textit{B}=3K$, $E_{c}=2.32K$. (b), (d) Single-terminal tunneling conductance as a function of bias voltage $V$ and Zeeman energy $V_{Z}$ of the proximitized nanowire, with MZM (b) and ABS (d) at the wire end respectively. The splittings at $V_{Z}=3meV$ are labeled as $\epsilon_{0}$ and $\epsilon_{A}$.}
\end{figure}

Finally, we consider the region (\rom{4}) corresponding to a trivial SC phase with $\Delta_\textit{B}>E_{c}$. In this region, the island at any $V_g$ is a pool of Cooper pairs at BCS ground state. Then the conductance peak arises only at the degenerate point of adding two electrons (e.g. electron number from $N$ to $N+2$) to the nanowire, because single electron tunneling process is blockaded by the SC gap. In this case, the conductance near the CB peaks due to the sequential tunneling of Cooper pairs is (see SI~\cite{supp} for more details)
\begin{align}
G_{2e} =& \frac{\left(2e\right)^{2}}{h}\frac{g_{l}^{2}g_{r}^{2}}{g_{l}^{2}+g_{r}^{2}}\frac{A}{T}W_{eq}\!\left(N\right)\!\int\!\! d\xi_{p_1}\!\! \int\!\! d\xi_{p_2}\!\! \int\!\! dE f\!\left(\xi_{p_1}\right)  \nonumber\\
&\times  f\!\left(\xi_{p_2}\right) \delta \left(E\!+\!U_{N+2}\!-\!U_{N}\!-\!\xi_{p_1}\!-\!\xi_{p_2}\right)P\left(E\right),\label{eq:general Cooper pair CB peak}
\end{align}
which describes the Andreev reflection process that a lead-electron in the level $p_1$  jumps into the nanowire, and reflects a hole to the level $p_2$ in the lead. The energy conservation requires $\xi_{p_1}+\xi_{p_2}+U(N)=U(N+2)+E$ where $E$ is the energy exchange with the environment. Near the peak, $W_{\mathrm{eq}}\left(N\right)\approx\left(1+e^{-\left(U_{N+2}-U_{N}\right)/T}\right)^{-1}$. Because the Andreev reflection is a second-order process, the lead-island coupling becomes: $\Gamma^{2e}=Ag^{2}T/2\pi$ ($A\sim 1$ is a dimensionless number representing the amplitude for the Andreev reflection~\cite{supp}). The transport is incoherent under the condition $\Gamma^{2e}\ll T$ which is is equivalent to saying $g^2\ll 1$.
The dissipative CB conductance of Cooper
pairs at $r=0.5$ is shown in Fig.~\ref{G_Splitting}(c). The peak height is also suppressed as
$T$ decreases to zero.

{\em Conclusion and discussion--}. So far, we have shown that under the condition $g\delta \ll T \lesssim g \Delta_\textit{Topo}$, $g^2\ll 1$, all the states in trivial phases carry incoherent transport, which will be suppressed as reducing the temperature in the presence of a dissipative environment. While the Majorana teleportation will maintain coherence, and the conductance will increase when the temperature goes down. This behavior could be a clear benchmark to determine the coherent nature of Majorana devices.

In addition, we emphasize that this two-terminal transport experiment for certain conditions can also tell the MZM from other contaminations like Andreev bound states (ABS), especially the zero-energy ABS (or equivalently two decoupled quasi-Majorana state) which can arise from smooth potential at wire end~\cite{kells2012PRB,CXLiuABSMZM,moore2018twoTerminal,VuikQM2018,moore2018quantized}. The single-terminal tunneling conductance of such devices can also have a quantized zero-bias peak~\cite{moore2018twoTerminal} which could mimic the MZM. By adjusting the 
potential of one end~\cite{supp}, the two quasi-Majorana components can have finite coupling regardless of the situation (either MZM or ABS) of the other end, which will generate the splitting in the single-terminal tunneling conductance as shown in Fig.~\ref{G_Splitting} (d), similar to the splitting of Majorana case (Fig.~\ref{G_Splitting} (b)). Denoting the splitting of MZM and ABS as $\epsilon_0$ and $\epsilon_A$, when $T\ll\epsilon_{0/A}$, the splittings in these two cases will lead to completely different two-terminal CB conductance signals.  In the Majorana case, the splitting comes from the coupling between the two Majorana modes at two ends, which will only shift the position of the CB peak by $\epsilon_0$ as shown in Fig.~\ref{G_Splitting} (a). In the ABS case, the splitting comes from the coupling between two quasi-Majorana components of ABS located at the same end; therefore, the two quasi-Majorana modes form a localized fermionic state~\cite{supp}. Then the non-local teleportation disappears, while only the CB peaks caused by the Cooper-pair tunneling (2-e) exist as shown in Fig.~\ref{G_Splitting} (c). In this case, the single electron tunneling needs the electron to go through the wire via the localized ABS. Because the coupling of the  ABS to the near lead is big and to the far lead is exponentially small, this is a highly asymmetric resonant tunneling process. The resulting conductance is quite small and according to the results in~\cite{Dong2012Nature,DongPRB2014}, dissipation can significantly suppress the conductance peak, i.e. showing opposite temperature dependence compared to the true Majorana case.

\begin{acknowledgments}
The authors acknowledge the support from Thousand-Young-Talent program of China, and the startup grant from State Key Laboratory of Low-Dimensional Quantum Physics and Tsinghua University. H.Z. also acknowledges the support from Beijing Academy of Quantum Information Sciences.
\end{acknowledgments}

\appendix

\begin{widetext}

\section*{Supplementary Material for\\``Reveal the non-local coherent nature from a dissipative Majorana teleportation''}
In this supplementary material, we will provide some details about: ~\ref{section01}.) Estimations of experimental parameters,~\ref{section02}.) the dissipative single electron tunneling conductance near the CB peak,~\ref{section03}.) the dissipative Cooper pair tunneling conductance near the CB peak,~\ref{section04}.) Distinguish between Majorana and ABS from two-terminal transport.

\section{Estimations of experimental parameters.\label{section01}}

The equivalent circuit diagram of the set-up is shown in Fig.~1(d) of the main text.
Two junctions on both sides of the nanowire are characterized by capacitance
$C_{j}$ and conductance $g_{j}$. The total capacitance between the
nanowire and the ground is $C_{\Sigma}=C_{l}+C_{r}+C_{g}$ which determines
the charging energy $E_{c}=e^{2}/2C_{\Sigma}$. While the capacitance
of the RC circuit is $C=C_{l}C_{r}/\left(C_{l}+C_{r}\right)$. The total resistance of the RC circuit is $R=R_l+R_r$, and the dissipation strength can be described by a dimensionless parameter $r=R/R_K$ where $R_K=h/e^2$ is the von-Klitzing resistance. For the convenience of the discussion, the capacitances of two junctions are taken the same with $C_{l}=C_{r}$ (e.g.  choose $2\times10^{-16}\text{F}$ for the energy scale comparison) and $C_{g}$ can be small $C_{g}\ll C_{l,r}$; for the general capacitance cases, our main conclusion is still valid~\cite{DongPRB2014}. Then the charging energy are of
the value $E_{c}\approx e^{2}/\left[\left(C_{l}+C_{r}\right)\right]=2.32K$
and $\omega_{R}=\left(RC\right)^{-1}=3r^{-1}K$.
With $r$ ranging from $0.5$ to $1.9$ and temperature $T$ ranging from $0.02K$ to $0.12K$ in the main text. The induced SC gap $\Delta_B$ is chosen such that: $\Delta_B>E_c\gg T$ in region (\rom{4}), $E_c>\Delta_B\gg T$ in region (\rom{3}), and gap-closing ($\Delta_B$ becomes the smallest energy scale) in region (\rom{2}). In principle, all the parameters are experimentally achievable and meet the conditions of the energy scales we study.

\section{The dissipative single electron tunneling conductance near the CB peak.\label{section02}}

The nanowire can be treated as a quantum dot shown in Fig.~\ref{FigDot}. The quantum dot and two leads are connected through the left and right barrier potentials. The energy levels of the quantum dot includes discrete levels with level spacing $\delta$ and an induced SC gap. The Fermi seas for the leads are shown in the red region, where the chemical potential of the left (right) lead is zero ($eV$ by the bias voltage). When the induced SC gap is small ($\Delta_B<E_c$), the tunneling current also includes the contribution from the single electron sequential tunneling processes. In this case, the current through the left barrier (which is the same as that through the right) is given by
\begin{align}
I & =-e\sum_{\alpha}\sum_{\left\{ n_{dot}\right\} }W\left(\left\{ n_{dot}\right\} \right)\left(\delta_{n_{\alpha},0}\Gamma_{0\rightarrow\alpha}^{l}-\delta_{n_{\alpha},1}\Gamma_{\alpha\rightarrow0}^{l}\right)\label{eq:I_original01}
\end{align}
where $\Gamma_{0\rightarrow\alpha}^{j}$ is the tunneling rate for the cases where an electron tunnels from the lead$-j$ to the state $\left|\alpha\right\rangle$ of the dot, and $\Gamma_{\alpha\rightarrow 0}^{j}$ is the tunneling rate of the reverse process. $n_{\alpha}$ (0 or 1) is the occupation number of the dot state $\alpha$ , and $W\left(\left\{ n_{dot}\right\} \right)$ is the probability of the occupation $\left\{ n_{dot}\right\} =\left\{ n_{1},n_{2},\ldots\right\}$. The electron tunneling from leads to a state $\left|\alpha\right\rangle$ in the dot and from $\left|\alpha\right\rangle$ to the leads change the probability $W\left(\left\{ n_{dot}\right\} \right)$. The changing rate should satisfy the master equation:
\begin{align}
\frac{\partial W\left(\left\{ n_{dot}\right\} \right)}{\partial t}
=&-\sum_{\alpha}  W\left(\left\{ n_{dot}\right\}\right)\delta_{n_{\alpha},0}
\left(\Gamma_{0\rightarrow\alpha}^{l}+\Gamma_{0\rightarrow\alpha}^{r}\right)
-\sum_{\alpha}  W\left(\left\{ n_{dot}\right\} \right)\delta_{n_{\alpha},1}
\left(\Gamma_{\alpha\rightarrow 0}^{l}+\Gamma_{\alpha\rightarrow 0}^{r}\right) \nonumber\\
&+\sum_{\alpha}  W\left(n_{1},\ldots,n_{\alpha-1},1,n_{\alpha+1},\ldots\right)\delta_{n_{\alpha},0}
 \left(\Gamma_{\alpha\rightarrow 0}^{l}+\Gamma_{\alpha\rightarrow 0}^{r}\right) \nonumber\\
&+\sum_{\alpha}  W\left(n_{1},\ldots,n_{\alpha-1},0,n_{\alpha+1},\ldots\right)\delta_{n_{\alpha},1}
 \left(\Gamma_{0\rightarrow\alpha}^{l}+\Gamma_{0\rightarrow\alpha}^{r}\right).\label{eq:changingrateofW}
\end{align}
Subsequently, we look at how to compute tunneling rate $\Gamma_{\alpha 0}^{j}$ for our model, solve the probability $W\left(\left\{ n_{dot}\right\} \right)$, and obtain the conductance formula.

The total Hamiltonian of the system is shown in Eq.~(2) of the main text.
The states of the leads are labeled by $p$ and the states of the quantum dot are labeled
by $\alpha$. The tunneling Hamiltonian is
$H_{\textit{T}}=\sum_{j,\sigma}t_{j}d_{\sigma}^{\dagger}\left(\bold{r}_j\right)c_{\sigma}\left(\bold{r}_j\right)e^{-i\varphi_{j}}+\textit{H.c.}$, where $d_{\sigma}\left(\bold{r}_j\right)=L^{\frac
{1}{2}}\sum_{\alpha}u_{\alpha}\left(\bold{r}_j,\sigma\right)\gamma_{\alpha}+v_{\alpha}^{*}\left(\bold{r}_j,\sigma\right)\gamma_{\alpha}^{\dagger}$ and $c_{\sigma}\left(\bold{r}_j\right)=\Omega_{j}^{\frac
{1}{2}}\sum_{p,\sigma}\phi_{p}\left(\bold{r}_j\right)c_{p\sigma}$ are the electron
field operators in the nanowire and the lead respectively, with $u_{\alpha}/v_{\alpha}\left(\bold{r}_j,\sigma\right)$ and $\phi_{p}\left(\bold{r}\right)$ being the wavefunctions
of BDG Hamiltonian and the wavefunction of the lead-electrons respectively, where $\sigma=\uparrow,\downarrow$ is the spin index and $j=l,r$ is the junction index. $L$ is the length of the nanowire and $\Omega_j$ is the volume of lead$-j$. Treating $H_{\textit{T}}$ as perturbation we can obtain the tunneling rate
\begin{align}
\Gamma_{0\rightarrow\alpha}^{j} & =\frac{g_{j}\Lambda\delta}{h}\int\mathrm{d}\xi_{p} f\left(\xi_{p}-\mu_{j}\right)P\left(-U_{N+1}+U_{N}-\epsilon_{\alpha}+\xi_{p}\right),\label{eq:Gamma0alpha}\\
\Gamma_{\alpha\rightarrow0}^{j} & =\frac{g_{j}\Lambda\delta}{h}\int\mathrm{d}\xi_{p} \left[1-f\left(\xi_{p}-\mu_{j}\right)\right]P\left(U_{N}-U_{N-1}+\epsilon_{\alpha}-\xi_{p}\right),\label{eq:Gammaalpha0}
\end{align}

\noindent where $f\left(x\right) = \left(1+e^{x/T}\right) ^{-1}$
is the Fermi-Dirac distribution. $\mu_{j}$ is the
chemical potential of the lead-$j$ and we have set $\mu_{l}=0$,
$\mu_{r}=eV$ as show in Fig.~\ref{FigDot}.  The barrier transmission is proportional to the dimensionless conductance $g_{j}=4\pi^{2}\left|t_{j}\right|^{2}\nu_{j}\nu$,
where $\nu_{j}$ is the density of states (DOS) of the lead-$j$ and $\nu=\delta^{-1}$
is the DOS of the nanowire when the SC gap closes ($\Delta_{B}=0$). $\Lambda$ is related to the SC-nanowire wavefunction at the tunneling location: $\Lambda=L\sum_{\sigma}\left|u_{\alpha}\left(\bold{r}_{j},\sigma\right)\right|^{2}$. Because the temperature is much smaller than the gap of the s-wave SC shell, almost the entire wavefunction $u_{\alpha}\left(\bold{r},\sigma\right)$ that contribute to the transport is located inside the nanowire and thus $\Lambda\sim 1$ which is treated as a constant value in region (\rom{2}) and (\rom{3}) of Fig.~1(e) in the main text~\cite{van2016PRB}. 

The function $P\left(E\right)=\left(2\pi\hbar\right)^{-1}\int_{-\infty}^{+\infty}\mathrm{d}t\left\langle e^{i\varphi_j(t)}e^{-i\varphi_j(0)}\right\rangle e^{iEt/\hbar}$ which can be interpreted as the probability of exchanging energy $E$ with the environment. They follow the standard relation $P\left(-E\right)=P\left(E\right)e^{-E/T}$, which means the ratio between the probability to emit energy into the environment and the probability to absorb energy from the environment is a Boltzmann factor. In the absence of dissipation, we have $P\left(E\right)=\delta\left(E\right)$ and Eq.~(\ref{eq:Gamma0alpha}) becomes the familiar tunneling rate \cite{glazman2005review,van2016PRB}. With $C_l=C_r$, $\left\langle e^{i\varphi_j(t)}e^{-i\varphi_j(0)}\right\rangle=\left\langle e^{i\varphi(t)/2}e^{-i\varphi(0)/2}\right\rangle$ where the phase $\varphi=\varphi_l-\varphi_r$ corresponds to the charge transfer between the left and right junction. For ohmic dissipation and $\omega_R/T\gg 1$, we can evaluate the integral
\begin{align}
P\left(E\right) =
\frac{\left(2\pi\right)^{\frac{r}{2}-2}}{\omega_{R}\mathit{\Gamma}(\frac{r}{2})}\left(\frac{ T}{\omega_{R}}\right)^{\frac{r}{2}-1}e^{\frac{E}{2T}}\left|\mathit{\Gamma}\left(\frac{r}{4}+i\frac{E}{2\pi T}\right)\right|^{2}
,\label{eq:PEsupp01}
\end{align}
where $\mathit{\Gamma}\left(z\right)$ is the gamma function.

\begin{figure}
\centering
\includegraphics[width=0.6\columnwidth]{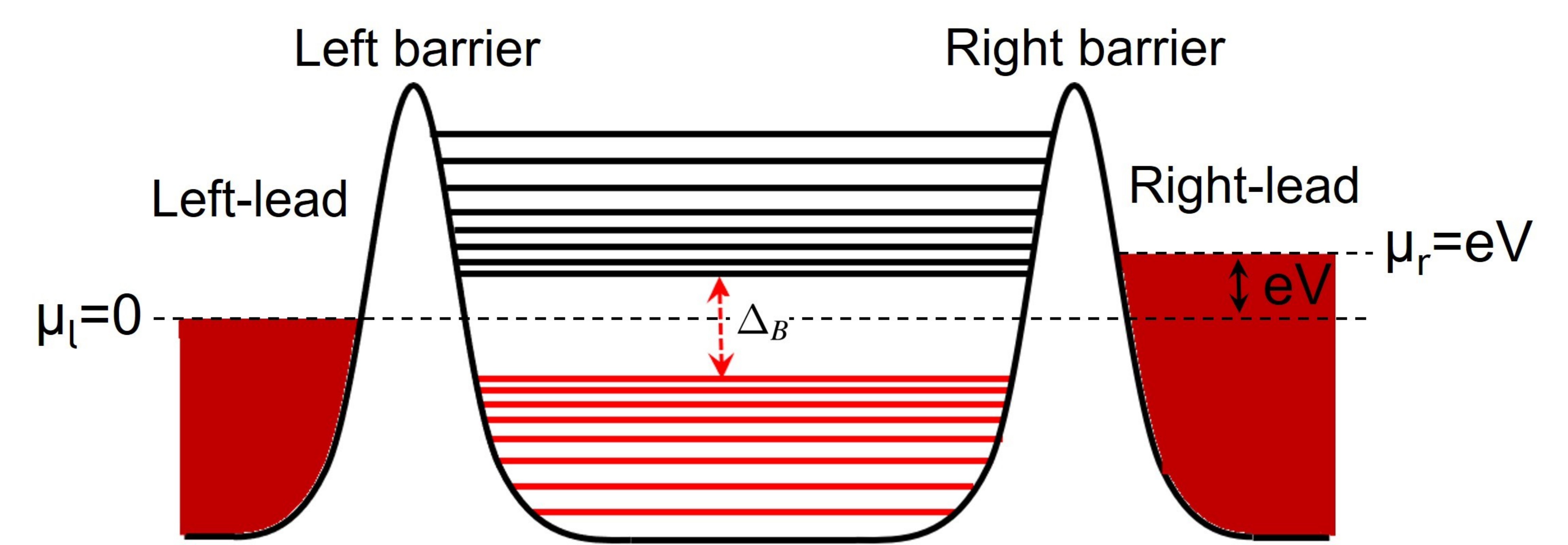}
\caption{\label{FigDot}Illustration of the energy levels of the proximitized nanowire and the Fermi seas of the leads.}
\end{figure}

To solve the master equation for probability function, we assume an expansion form for $W\left(\left\{ n_{dot}\right\} \right)$ in order of bias voltage $V$, and only keep the linear order terms
\begin{equation}
W\left(\left\{ n_{dot}\right\} \right)=W_{eq}\left(\left\{ n_{dot}\right\} \right)\left(1+\frac{eV}{T}\Psi\left(\left\{ n_{dot}\right\} \right)\right),\label{eq:linearW}
\end{equation}
where $W_{eq}\left(\left\{ n_{dot}\right\} \right)$ is the probability distribution
in equilibrium when no bias voltage is applied and has the form $W_{eq}\left(\left\{ n_{dot}\right\} \right)=Z^{-1}\exp\left[-\left(\sum_{\alpha}\epsilon_{\alpha}n_{\alpha}+U_{N}\right)/T\right]$ where $Z=\sum_{\left\{ n_{dot}\right\} }\exp\left[-\left(\sum_{\alpha}\epsilon_{\alpha}n_{\alpha}+U_{N}\right)/T\right]$. The steady state occupation probabilities can be determined by setting
$\partial W\left(\left\{ n_{dot}\right\} \right)/\partial t=0$. Then for each $\alpha$ in Eq.~(\ref{eq:changingrateofW}), we can obtain

\begin{align}
\Psi\left(n_{1},\ldots,n_{\alpha-1},1,n_{\alpha+1},\ldots\right)=\Psi\left(n_{1},\ldots,n_{\alpha-1},0,n_{\alpha+1},\ldots\right)+\frac{g_{r}}{g_{r}+g_{l}}.\label{Psirelation}
\end{align}

\noindent Combining Eq.~(\ref{eq:I_original01}), (\ref{eq:linearW}) and (\ref{Psirelation}) and keeping only the leading terms in $V$, we obtain the zero-bias conductance:

\begin{equation}
G=\frac{e^{2}}{h}\frac{g_{l}g_{r}}{g_{r}+g_{l}}\frac{\Lambda\delta}{T}\sum_{N}\sum_{\alpha}W_{eq}\left(N\right)\left[1-F_{\mathrm{eq}}\left(\epsilon_{\alpha}|N\right)\right]\left[\int dEP\left(E\right)f\left(E+\epsilon_{\alpha}+U_{N+1}-U_{N}\right)\right],\label{eq:single electron G}
\end{equation}

\noindent where $W_{eq}\left(N\right)=\sum_{\left\{ n_{dot}\right\} }W_{eq}\left(\left\{ n_{dot}\right\} \right)\delta_{N,\sum_{\alpha}n_{\alpha}}$
is the probability that the isolated dot contains $N$ electrons
in equilibrium, and $F_{\mathrm{eq}}\left(\epsilon_{\alpha}|N\right)=W^{-1}_{\mathrm{eq}}(N)\sum_{\left\{ n_{dot}\right\} }P_{\mathrm{eq}}\left(\left\{ n_{dot}\right\} \right)\delta_{n_{\alpha},1}\delta_{N,\Sigma_{\alpha}n_{\alpha}}$
is the equilibrium conditional probability that level $\alpha$ is
occupied given the dot electron number $N$. In the main text, we only consider the limit $T\ll E_{c}$ and the region
near the CB peak. In this limit, we can neglect the summation over $N$ in Eq.~(\ref{eq:single electron G}).

\subsection{The normal metal regime.}
Near the SC gap closing point in region (\rom{2}) of Fig.~1(e) of the main text,
the nanowire corresponds to a metallic phase and behaves like a normal
metal island. In the case $\delta\ll T$, the discrete energy spectrum
in the island can be treated as a continuum. In this limit, $F_{\mathrm{eq}}\left(\epsilon_{\alpha}|N\right)$
can be approximated as Fermi-Dirac distribution $f\left(\epsilon_{\alpha}-\mu\left(N\right)\right)$
and $\mu\left(N\right)$ can be determined by $\sum_{\alpha}f\left(\epsilon_{\alpha}-\mu\left(N\right)\right)=N$, and choose $\mu\left(N\right)\approx\mu=0$. Near the CB peak, a lead-electron can tunnel into a dot state near the Fermi surface; then the energy change of the dot is $U_{N+1}-U_N$. Therefore, including both $N$ and $N+1$ electron state, the probability of $N$ electrons is
\begin{align}
W_{\mathrm{eq}}\left(N\right)=\frac{1}{1+e^{-\left(U_{N+1}-U_{N}\right)/T}}.
\end{align}
The terms beyond $U_{N}$ and $U_{N+1}$ are neglected for $T\ll E_{c}$.
Performing the integration (the summation over $\alpha$) we can obtain the conductance
in the metal phase:
\begin{equation}
G=\frac{\Lambda e^{2}}{h}\frac{g_{l}g_{r}}{g_{r}+g_{l}}\frac{1
}{1+e^{-\left(U_{N+1}-U_{N}\right)/T}}\left[\int dEP\left(E\right)\frac{\left(E+U_{N+1}-U_{N}\right)/T}{e^{\left(E+U_{N+1}-U_{N}\right)/T}-1}\right].\label{eq:metal G}
\end{equation}
The results shown in Fig.~2(b) of the main text are obtained numerically from evaluating Eq.~(\ref{eq:metal G}). The CB peaks appear at $U_{N+1}-U_N=0$, from which we can obtain the peak conductance:
\begin{equation}
G^\textit{peak}=\frac{\Lambda e^{2}}{h}\frac{g_{l}g_{r}}{g_{r}+g_{l}}\frac{\left(2\pi\right)^{\frac{r}{2}-2}}{\mathit{\Gamma}(\frac{r}{2})}\left(\frac{ T}{\omega_{R}}\right)^{\frac{r}{2}}\left[\int ds\left|\mathit{\Gamma}\left(\frac{r}{4}+i\frac{s}{\pi}\right)\right|^{2}\frac{s}{\sinh s}\right]\propto T^{\frac{r}{2}}.\label{eq:metal Gpeak}
\end{equation}

\subsection{The single electron tunneling regime with $T\ll \Delta_B$.}

In the region (\rom{3}) with the limit $T\ll \Delta_B$, the SC gap
is big enough to prohibit thermal excitations of quasiparticle states, and therefore the equilibrium conditional probability for level $\alpha$ simply becomes $F_{\mathrm{eq}}\left(\epsilon_{\alpha}|N\right)=0$. Near the CB peak, a lead-electron can tunnel into a dot quasiparticle state with energy $\epsilon_{\alpha}$; then the energy change of the dot is $U_{N+1}+\epsilon_{\alpha}-U_N$. Therefore, including both $N$ and $N+1$ electron state, the probability of $N$ electrons is

\begin{align}
W_{\mathrm{eq}}\left(N\right)=\frac{1}{1+\sum_{\alpha}e^{-\left(U_{N+1}+\epsilon_{\alpha}-U_{N}\right)/T}},\label{eq:W_single_e}
\end{align}

\noindent Feeding Eq.~(\ref{eq:W_single_e}) into Eq.~(\ref{eq:single electron G}), we can obtain the conductance:

\begin{equation}
G=\frac{\Lambda e^{2}}{h}\frac{\delta}{T}\frac{g_{l}g_{r}}{g_{r}+g_{l}}\frac{1}{1+\sum_{\alpha}e^{-\left(\epsilon_{\alpha}+U_{N+1}-U_{N}\right)/T}}\sum_{\alpha}\int dEP\left(E\right)\frac{1}{1+e^{\left(E+\epsilon_{\alpha}+U_{N+1}-U_{N}\right)/T}}.\label{eq:single electron G-1}
\end{equation}

\noindent The results shown in Fig.~{\ref{Single_e_EtaB}(a)} below and Fig.~3(c) of the main text are obtained by numerically evaluating Eq.~(\ref{eq:single electron G-1}). Because the states with energy $\epsilon_{\alpha}$ close to $\Delta_B$ contribute the most to the conductance, the peak positions will be shifted by about $\Delta_B/\left(2E_c\right)$. Besides, the conductance shows a special temperature scaling $T^{r/2-\eta_B}$, where an extra scaling factor $\eta_B$ tends to one for $\Delta_\textit{B}\gg T$ and zero at the gap closing point.

\begin{figure}
\centering
\includegraphics[width=1\columnwidth]{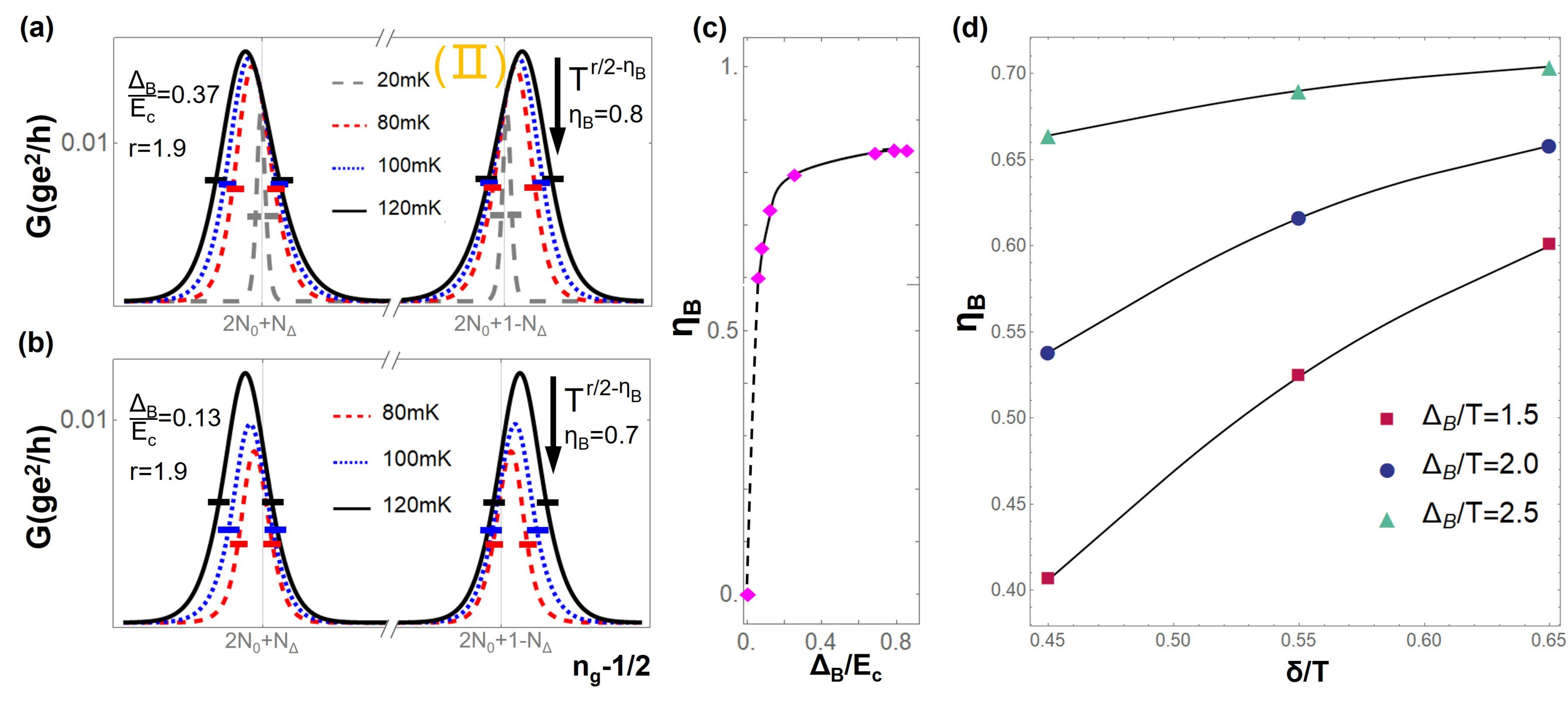}
\caption{\label{Single_e_EtaB} (a) The dissipative CB conductance of a superconductor in region ({\rom{3}}) of Fig.~1(e) where $T\ll\Delta_\textit{B}<E_{c}$ with $\Delta_\textit{B}=0.86K$, $E_{c}=2.32K$. The peak positions shift about $N_{\Delta}$ with $N_{\Delta}\equiv \Delta_B/\left(2E_c\right)$. (b) The dissipative CB conductance in the region between (\rom{2}) and (\rom{3}) when $T\lesssim\Delta_\textit{B}<E_c$ with $\Delta_\textit{B}=0.3K$, $E_{c}=2.32K$. (c) The relation between the scaling factor $\eta_B$ in $G\propto T^{\frac{r}{2}-\eta_B}$ and $\Delta_\textit{B}$ with $E_c=2.32K$ and $T=0.1K$. The calculated data is represented by magenta diamonds. (d) The relation between $\eta_B$ and the level spacing $\delta$ for three different relatively small $\Delta_B$ with $T=0.1K$, $E_{c}=2.32K$.}
\end{figure}

\subsection{The regime $T\lesssim\Delta_{B}$.\label{regimeTsimD}}

To find the behavior of the scaling factor $\eta_B$ we consider the crossover between single electron (still SC) regime (\rom{3}) and a metallic (non-SC) regime (\rom{2}). Between (\rom{2}) and (\rom{3}), $T<\Delta_{B}$ but not in the limit $T\ll\Delta_B$, the quasiparticle states could be thermally excited which can modify the transport signatures. In this phase, the quasiparticle number is not conserved, only its parity is conserved, the total number of excited states will have
the same parity with $N$. When $N$ is even, we have

\begin{align}
W_{\mathrm{eq}}(N)=\frac{\sum_{\left\{ n_{dot}\right\} }\delta_{even,\sum n_{\alpha}}\exp\left[-\frac{1}{T}\left(\sum_{\alpha}\epsilon_{\alpha}n_{\alpha}+U_{N}\right)\right]}{\sum_{\left\{ n_{dot}\right\} }\delta_{even,\sum n_{\alpha}}\exp\left[-\frac{1}{T}\left(\sum_{\alpha}\epsilon_{\alpha}n_{\alpha}+U_{N}\right)\right]+\sum_{\left\{ n_{dot}\right\} }\delta_{odd,\sum n_{\alpha}}\exp\left[-\frac{1}{T}\left(\sum_{\alpha}\epsilon_{\alpha}n_{\alpha}+U_{N+1}\right)\right]},
\end{align}
\begin{align}
1-F_{\mathrm{eq}}\left(\epsilon_{\alpha}|N\right)=\frac{\sum_{\left\{ n_{i\neq\alpha}\right\} }\delta_{even,\sum n_{\alpha}}\exp\left[-\frac{1}{T}\left(\sum_{\alpha}\epsilon_{\alpha}n_{\alpha}\right)\right]}{\sum_{\left\{ n_{dot}\right\} }\delta_{even,\sum n_{\alpha}}\exp\left[-\frac{1}{T}\left(\sum_{\alpha}\epsilon_{\alpha}n_{\alpha}\right)\right]}.
\end{align}
Substituting these equations into Eq.~(\ref{eq:single electron G}), we obtain
\begin{align}
G=  \frac{\Lambda e^{2}}{h}\frac{\delta}{T}\frac{g_{l}g_{r}}{g_{r}+g_{l}}\frac{1}{F_{even}+F_{odd}e^{-\left(U_{N+1}-U_{N}\right)/T}}\sum_{\alpha}D_{even}\left(\epsilon_{\alpha}\right)\int dE P\left(E\right)\frac{1}{1+e^{\left(E+\epsilon_{\alpha}+U_{N+1}-U_{N}\right)/T}},\label{eq:G_even}
\end{align}
where
\begin{align}
F_{even/odd}=&\sum_{\left\{ n_{dot}\right\} }\delta_{\sum n_{\alpha},even/odd}\exp\left[-\frac{1}{T}\left(\sum_{\alpha}\epsilon_{\alpha}n_{\alpha}\right)\right],\\
D_{even/odd}\left(\epsilon_{\alpha}\right)=&\sum_{\left\{ n_{i\neq\alpha}\right\} }\delta_{\sum n_{\alpha},even/odd}\exp\left[-\frac{1}{T}\left(\sum_{\alpha}\epsilon_{\alpha}n_{\alpha}\right)\right].
\end{align}
When $N$ is odd,
\begin{align}
W_{\mathrm{eq}}(N)=\frac{\sum_{\left\{ n_{dot}\right\} }\delta_{odd,\sum n_{\alpha}}\exp\left[-\frac{1}{T}\left(\sum_{\alpha}\epsilon_{\alpha}n_{\alpha}+U_{N}\right)\right]}{\sum_{\left\{ n_{dot}\right\} }\delta_{odd,\sum n_{\alpha}}\exp\left[-\frac{1}{T}\left(\sum_{\alpha}\epsilon_{\alpha}n_{\alpha}+U_{N}\right)\right]+\sum_{\left\{ n_{dot}\right\} }\delta_{even,\sum n_{\alpha}}\exp\left[-\frac{1}{T}\left(\sum_{\alpha}\epsilon_{\alpha}n_{\alpha}+U_{N+1}\right)\right]},
\end{align}
\begin{align}
1-F_{\mathrm{eq}}\left(\epsilon_{\alpha}|N\right)=\frac{\sum_{\left\{ n_{i\neq\alpha}\right\} }\delta_{odd,\sum n_{\alpha}}\exp\left[-\frac{1}{T}\left(\sum_{\alpha}\epsilon_{\alpha}n_{\alpha}\right)\right]}{\sum_{\left\{ n_{dot}\right\} }\delta_{odd,\sum n_{\alpha}}\exp\left[-\frac{1}{T}\left(\sum_{\alpha}\epsilon_{\alpha}n_{\alpha}\right)\right]}.
\end{align}
Substituting the upper two formulas into Eq.~(\ref{eq:single electron G})
\begin{align}
G=\frac{\Lambda e^{2}}{h}\frac{\delta}{T}\frac{g_{l}g_{r}}{g_{r}+g_{l}}\frac{1}{F_{odd}+F_{even}e^{-\left(U_{N+1}-U_{N}\right)/T}}\sum_{\alpha}D_{odd}\left(\epsilon_{\alpha}\right)\int dE P\left(E\right)\frac{1}{1+e^{\left(E+\epsilon_{\alpha}+U_{N+1}-U_{N}\right)/T}}.\label{eq:G_odd}
\end{align}
By numerically evaluating Eq.~(\ref{eq:G_even}) and (\ref{eq:G_odd}), we obtain the CB conductance shown in Fig.~{\ref{Single_e_EtaB}}(b). Together with Eq.~(\ref{eq:single electron G-1}), we numerically plot the $\eta_B$ as a function of  SC gap $\Delta_B$, and find that $\eta_B$ decreases as the gap is reduced as shown in Fig.~\ref{Single_e_EtaB}(c). Note that we have $\eta_B=0$ at the gap closing point.

For this regime, we in principle need to consider contributions of all the levels above the SC gap $\Delta_B$. The level spacing $\delta$ above the gap is very small $\delta\ll T\lesssim \Delta_B$, which results in a large number of levels above the gap, and the exact numerical evaluation becomes intractable. In fact, we only want to qualitatively understand the behavior of the scaling factor $\eta_B$ in conductance scaling $G\propto T^{r/2-\eta_B}$ as shown in the main text, and look at how does $\eta_B$ change as we reduce the SC gap (or vary magnetic field). Therefore, we include a truncation and only consider a few energy levels above the gap, which needs a relatively big $\delta$. In the numerical evaluation of the scaling factor $\eta_B$ of the main text, we take the level spacing $\delta=0.065K$ and $T\sim 0.1K$. Since $\Delta\sim T$, it seems that the situation is quite different from the real case in which $T\gg \delta$. Here we emphasize that as we decrease $\delta$ toward more realistic situation, the scaling factor $\eta_B$ will decrease as shown in Fig.~\ref{Single_e_EtaB}(d), which means that with the same $r$, the CB peak conductance will decrease faster as temperature is reduced, when $\delta$ becomes smaller (toward the value of a more realistic case). So, more realistic situations with smaller level spacing can even strengthen our results (peak conductance for non-Majorana incoherent regimes is strongly suppressed as the temperature is reduced in the presence of dissipation).

\section{The dissipative Cooper pair tunneling conductance near the CB peak.\label{section03}}
In region (\rom{4}), $\Delta_{B}>E_{c}$ and the CB
conductance exhibits oscillations with $2e-$ periodicity. Near the
CB peak, the transport is dominated by the Cooper pair tunneling or
the Andreev reflection process, while the single electron tunneling is suppressed
by the SC gap. In this case, the current through the left barrier is given by

\begin{align}
I_{2e}=-2e\left(W_{0}\Gamma_{0\rightarrow2}^{l}-W_{2}\Gamma_{2\rightarrow0}^{l}\right),\label{eq:I2e}
\end{align}
where $\Gamma_{0\rightarrow2}^{j}$ is the tunneling rate of an Andreev reflection process, which changes the number of electrons in the dot by two by adding a Cooper pair from lead-$j$ and $\Gamma_{2\rightarrow 0}^{j}$ is the tunneling rate of the reverse process which removes a Cooper pair from the dot to lead-$j$. $W_2$ and $W_0$ are the probabilities that the dot has and doesn't have an additional Cooper pair, respectively.
The tunneling of Cooper pairs into and from the dot will change the probabilities, and give us the standard master equation

\begin{align}
\frac{\partial W_{0}}{\partial t}=&-\left(\Gamma_{0\rightarrow2}^{l}+\Gamma_{0\rightarrow2}^{r}\right)W_{0}+\left(\Gamma_{2\rightarrow0}^{l}+\Gamma_{2\rightarrow0}^{r}\right)W_{2},\label{eq:W0}\\
\frac{\partial W_{2}}{\partial t}=&-\left(\Gamma_{2\rightarrow0}^{l}+\Gamma_{2\rightarrow0}^{r}\right)W_{2}+\left(\Gamma_{0\rightarrow2}^{l}+\Gamma_{0\rightarrow2}^{r}\right)W_{0}.\label{eq:W2}
\end{align}

\noindent From the second-order perturbation theory, we can obtain the Cooper pair tunneling rates:
\begin{align}
\Gamma_{0\rightarrow2}^{j}= & \frac{g_{j}^{2}A}{h}\int d\xi_{p_{1}}\int d\xi_{p_{2}}f\left(\xi_{p_{1}}-\mu_{j}\right)f\left(\xi_{p_{2}}-\mu_{j}\right)P\left(-U_{N+2}+U_{N}+\xi_{p_{1}}+\xi_{p_{2}}\right),\label{eq:Gamma02}\\
\Gamma_{2\rightarrow 0}^{j}=& \frac{g_{j}^{2}A}{h}\int d\xi_{p_{1}}\int d\xi_{p_{2}}\left[1-f\left(\xi_{p_{1}}-\mu_{j}\right)\right]\left[1-f\left(\xi_{p_{2}}-\mu_{j}\right)\right]P\left(U_{N+2}-U_{N}-\xi_{p_{1}}-\xi_{p_{2}}\right).\label{eq:Gamma20}
\end{align}
Note here $P\left(E\right)$ becomes
\begin{align}
P\left(E\right)=\frac{1}{2\pi\hbar}\int_{-\infty}^{+\infty}\mathrm{d}t\left\langle e^{i\varphi(t)}e^{-i\varphi(0)}\right\rangle e^{\frac{i}{\hbar}Et} =
\frac{\left(2\pi\right)^{2r-2}}{\omega_{R}\mathit{\Gamma}(2r)}\left(\frac{T}{\omega_{R}}\right)^{2r-1}e^{\frac{E}{2T}}\left|\mathit{\Gamma}\left(r+i\frac{E}{2\pi T}\right)\right|^{2}
.\label{eq:PEsupp}
\end{align}

\noindent In Eq.~(\ref{eq:Gamma20}) and (\ref{eq:Gamma02}), $A$ is a dimensionless number
representing the amplitude of the Andreev reflection:

\begin{align}
A=\frac{1}{\pi^{2}}\left|L\nu^{-1}\sum_{\alpha}
v_{\alpha}\left(\bold{r}_{j},\sigma\right)u_{\alpha}^{*}\left(\bold{r}_{j},-\sigma\right)\frac{1}{U_{N+1}-U_{N}+\epsilon_{\alpha}}\right|^{2},
\end{align}

For the steady state, we can set $\partial W_0/\partial t=0$ and $\partial W_2/\partial t=0$ in Eq.~(\ref{eq:W0}) and (\ref{eq:W2}). We also have the probability conservation condition $W_0+W_2=1$. After solving $W_0$ and $W_2$, we obtain the CB conductance due to the Cooper pair tunneling:
\begin{align}
G_{2e} & =\frac{\left(2e\right)^{2}}{h}\frac{g_{l}^{2}g_{r}^{2}}{g_{l}^{2}+g_{r}^{2}}\frac{A}{T}\frac{1}{1+e^{-\left(U_{N+2}-U_{N}\right)/T}}\int d\xi_{p_{1}}\int d\xi_{p_{2}}f\left(\xi_{p_{1}}\right)f\left(\xi_{p_{2}}\right)P\left(-U_{N+2}+U_{N}+\xi_{p_{1}}+\xi_{p_{2}}\right),\label{eq:G2egeneral}
\end{align}
which is the Eq.~(6) in the main text. The CB peaks appear at $U_{N+2}-U_N=0$, from which we can obtain the peak conductance:
\begin{equation}
G_{2e}^\textit{peak}=\frac{\left(2e\right)^{2}}{h}\frac{g_{l}^{2}g_{r}^{2}}{g_{l}^{2}+g_{r}^{2}}\frac{A\left(2\pi\right)^{2r-2}}{\mathit{\Gamma}(2r)}\left(\frac{T}{\omega_{R}}\right)^{2r}\left[\int ds\left|\mathit{\Gamma}\left(r+i\frac{s}{\pi}\right)\right|^{2}\frac{s}{\sinh s}\right]\propto T^{2r}.\label{2eGpeak}
\end{equation}

\begin{figure}[t]
\centering
\includegraphics[width=\columnwidth]{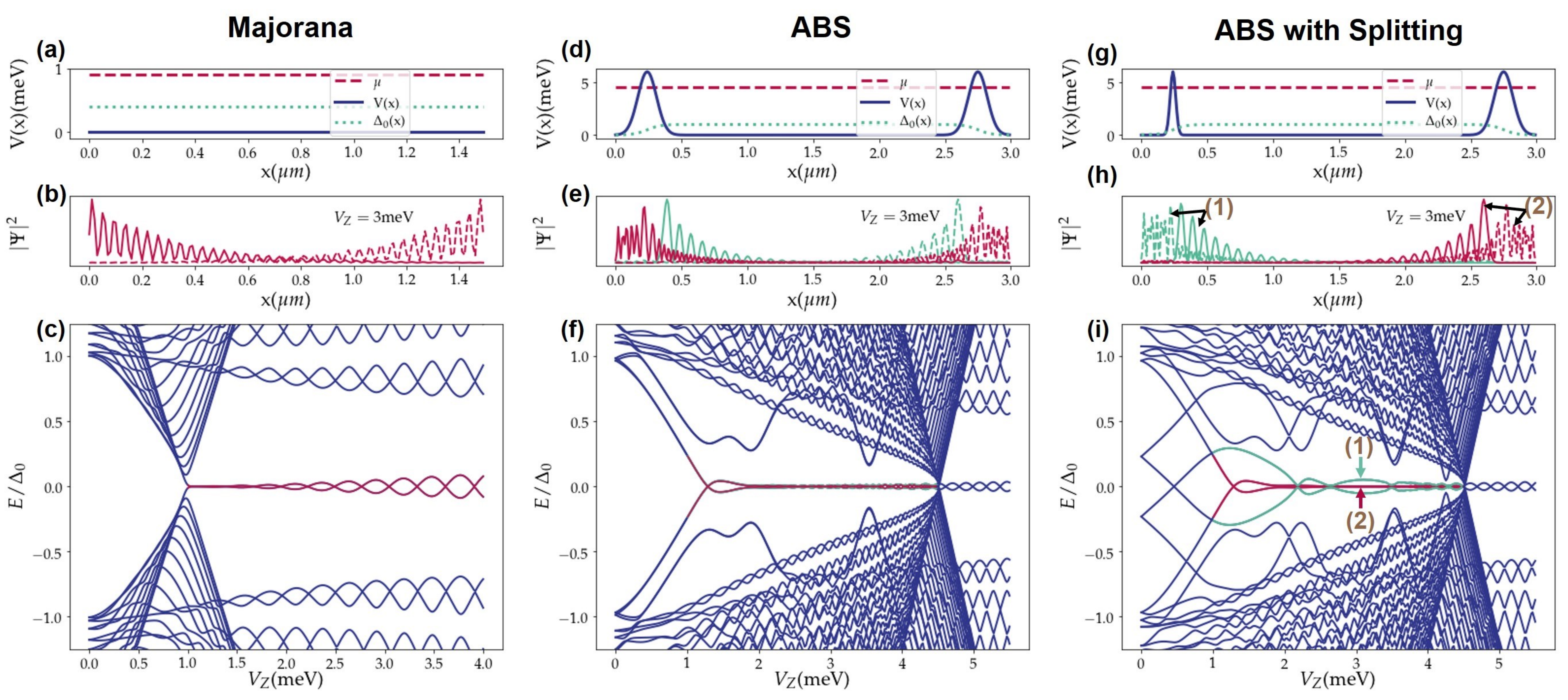}
\caption{\label{FigABS}(a),(d) and (g) are the potential profiles $V(x)$ for Majorana, ABS and ABS with splitting. The spatial distribution of chemical potential $\mu$ and SC gap under zero magnetic field $\Delta_0 (x)$ are also plotted in the picture. (c),(f) and (i) are the corresponding energy spectrum as a function of Zeeman field $V_Z$. (c) and (i) correspond to the single-terminal tunneling conductance spectrum shown in Fig. 3(b) and (d) in the main text. (b),(e) and (h) show the Majorana wavefuctions and the quasi-Majorana wavefuctions which make up the ABS in each case. The red line in (c) shows the MZM with splitting due to the finite size and the two Majorana modes at $V_Z =3meV$ are shown in (b). The red and green lines in (f) show that ABSs appear when there are smooth potentials at the wire ends. (i) shows that the zero-energy mode in (f) splits when sharpening the end potential. (h) shows that the two quasi-Majorana wavefunctions (green) of the split-mode (green line) in (f) are both localized at the left end.}
\end{figure}

\section{Distinguish between Majorana and ABS from two-terminal transport.\label{section04}}
Here we show that how the two-terminal transport experiment can tell the MZM from the near-zero-energy Andreev bound states (ABS). The zero-energy ABS can arise from smooth potential at wire end~\cite{kells2012PRB,CXLiuABSMZM,moore2018twoTerminal,VuikQM2018,moore2018quantized}. Fig.~\ref{FigABS}(f) shows the energy spectrum when there are smooth Gaussian-shape potentials at two wire ends (as shown in Fig.~\ref{FigABS}(d)). When the Zeeman field is big enough but smaller than the topological transition point $V_Z=4.52meV$, zero-energy ABSs show up at two wire ends. The single-terminal tunneling conductance taken at one end of such devices can also have a zero-bias peak which could mimic the MZM. The four decoupled quasi-Majorana components (whose definition can be found in~\cite{moore2018twoTerminal}) which make up two ABSs are plotted in Fig.~\ref{FigABS}(e) (two quasi-Majorana components make up one ABS). By making the potential sharper at one end (see Fig.~\ref{FigABS}(g)), one can couple the two quasi-Majorana states at that end and split the zero-energy spectral line (the green line in Fig.~\ref{FigABS}(i)) of the corresponding ABS. In the single-terminal tunneling experiment, this will only split the conductance peak (see Fig.~3(d) of the main text), which could also mimic the Majorana splitting due to the finite size (Fig.~3(b) of the main text). However since the two coupled quasi-Majorana states (green solid and dashed lines in Fig.~\ref{FigABS}(h)) are coupled and make up a localized fermionic ABS at the left end, which can no longer pair up with the states at the other end, the non-local teleportation via two quasi-Majorana states localized at different ends disappears. For two-terminal transport, a single electron can only tunnel through the whole wire via this fermionic ABS, which only strongly couples to one lead and exponentially decouples to the other. For such a highly asymmetric resonant tunneling, the conductance should be quite small; and in addition, the conductance will also be suppressed when decreasing the temperature in the case with dissipation~\cite{Dong2012Nature,DongPRB2014}. In the Majorana case as shown in Fig.~\ref{FigABS}(c), even if there is splitting, the two Majorana modes shown in Fig.~\ref{FigABS}(b) are still mostly localized at each end. The coherent Majorana teleportation still exists and the resonant peak height still increases approaching the quantized value when lowering the temperature in the case with dissipation. Only the resonant peak position will be shifted. In conclusion, by tuning the sharpness of the potential, the dissipative transport can distinguish the coherent Majorana teleportation with the incoherent ABS transport.

\end{widetext}

\bibliographystyle{apsrev4-1} 
\bibliography{DissipativeMaj}

\end{document}